\documentclass{article}
\usepackage{tabularx}
\usepackage{comment, arydshln, amsfonts}
\usepackage{kotex,spconf,amsmath,graphicx}
\usepackage{cite, url}
\usepackage{xcolor}
\usepackage{makecell}
\usepackage{multirow}
\usepackage[subtle]{savetrees}
\usepackage{hyperref}
\hypersetup{ 
    colorlinks=true,
    linkcolor=purple,
    filecolor=magenta,      
    urlcolor=purple,
}

\newcolumntype{C}{>{\small\centering\arraybackslash}X}

\title{Diff-SV: A Unified Hierarchical Framework for Noise-Robust Speaker Verification Using Score-Based Diffusion Probabilistic Models}
\name{Ju-ho Kim, Jungwoo Heo, Hyun-seo Shin, Chan-yeong Lim and Ha-Jin Yu$^\dag$\thanks{$^\dag$Corresponding author.}\thanks{This work was supported by the National Research Foundation of Korea(NRF) grant funded by the Korea government. (MSIT) (2023R1A2C1005744)}}
\address{School of Computer Science, University of Seoul}

\begin{document}
\ninept
\maketitle
\begin{abstract}
Background noise considerably reduces the accuracy and reliability of speaker verification (SV) systems. 
These challenges can be addressed using a speech enhancement system as a front-end module. 
Recently, diffusion probabilistic models (DPMs) have exhibited remarkable noise-compensation capabilities in the speech enhancement domain. 
Building on this success, we propose Diff-SV, a noise-robust SV framework that leverages DPM. 
Diff-SV unifies a DPM-based speech enhancement system with a speaker embedding extractor, and yields a discriminative and noise-tolerable speaker representation through a hierarchical structure. 
The proposed model was evaluated under both in-domain and out-of-domain noisy conditions using the VoxCeleb1 test set, an external noise source, and the VOiCES corpus. 
The obtained experimental results demonstrate that Diff-SV achieves state-of-the-art performance, outperforming recently proposed noise-robust SV systems. 
\end{abstract}

\section{Introduction} 
\label{sec:intro}
Speaker verification (SV) involves determining whether the speaker of a given utterance matches an authorized identity. 
Although recent advances in deep learning have yielded highly accurate deep neural network-based SV systems in clean and controlled environments \cite{variani2014deep, snyder2018x, desplanques2020ecapa, kim2022rawnext}, the performances of these systems degrade significantly under noisy, real-world conditions. 
Background noise diminishes speech intelligibility and quality, consequently hindering the extraction of accurate speaker representations \cite{wolfel2009distant, cai2020within}. 
Prior research has incorporated speech enhancement models as front-end modules in SV systems to mitigate the negative effects of noise \cite{plchot2016audio, novotny2019analysis, DBLP:conf/interspeech/KimHSY22}. 

In the speech enhancement field, generative models such as variational autoencoders \cite{leglaive2020recurrent}, generative adversarial networks \cite{soni2018time}, and flow-based models \cite{strauss2021flow} are widely used to generate clean speech from noisy speech. 
Recently, diffusion-based generative models have exhibited outstanding generation capabilities with respect to traditional generative models \cite{sohl2015deep, dhariwal2021diffusion}. 
As a representative approach, the denoising diffusion probabilistic model (DDPM) learns to predict Gaussian noise added to original data through a series of steps (i.e., forward diffusion process) and generates data by denoising random Gaussian noise iteratively via the Markov chain property using the trained model (i.e., reverse diffusion process) \cite{ho2020denoising}. 
Leveraging the DDPM's potent data generation capabilities, researchers have proposed speech enhancement systems that adopt diffusion principles. 
For instance, Lu et al. \cite{lu2021study} recovered clean speech from noisy speech using the DDPM's Markov chain process. 
Furthermore, CDiffuSE \cite{lu2022conditional} was developed for non-Gaussian real-world noise adaptation by incorporating noisy speech into the forward and reverse processes of DDPM. 

Despite its exceptional noisy-compensation capabilities, the direct application of a DDPM-based speech enhancement model may not be ideal for SV. 
Due to the stochastic generation process, DDPMs tend to produce low-consistency data (i.e., different data from the same input) \cite{ho2020denoising, DBLP:conf/iclr/SongME21}. 
Consequently, speaker embedding extractors may yield representations with increased and decreased intra- and inter-class variances, respectively, ultimately degrading SV performance.  
Furthermore, the numerous reverse steps owing to the Markov chain entail a considerable computational cost \cite{DBLP:conf/iclr/SongME21}, complicating the joint learning process with the SV system. 
These challenges can be alleviated using a score-based generation model \cite{song2020score} (refer to the Section \ref{sec:score} for details). 
In comparison with the stepwise prediction process of the DDPM, the score-based diffusion probabilistic model (DPM) generates data using fewer steps by numerically solving differential equations in continuous-condition sampling procedures. 
In addition, by employing an ordinary differential equation (ODE) solver instead of a stochastic differential equation (SDE) solver, deterministic data sampling can be achieved with minimal distribution perturbation. 

In this study, we propose a unified noise-robust SV framework named Diff-SV, which applies a score-based DPM as a front-end module of a speaker embedding extractor. 
Diff-SV uses not only a score-based DPM (denoiser) solely, but also an auxiliary enhancement system (enhancer) to remove noise from the input. 
Therefore, the denoiser only needs to remove residual noise contained in the features derived from the enhancer, allowing for highly stable and effective data sampling. 
Furthermore, the proposed framework combines the original noisy spectrogram, the enhanced feature from the enhancer, and the denoised feature from the denoiser to extract speaker embeddings. 
By simultaneously leveraging features with different characteristics, this hierarchical approach encourages embedding extraction that is robust to distributional perturbation and informative. 

VoxCeleb1 \cite{DBLP:conf/interspeech/NagraniCZ17} was used for model training, and this dataset was augmented with the MUSAN dataset \cite{snyder2015musan}. 
We evaluated the generalizability of the model  under in- and out-of-domain noise conditions using external data. 
Our proposed Diff-SV outperformed recent SV systems, demonstrating the efficacy of the framework through ablation experiments and visualizations. 

\begin{figure*}[!t]
    \includegraphics[width=\textwidth]{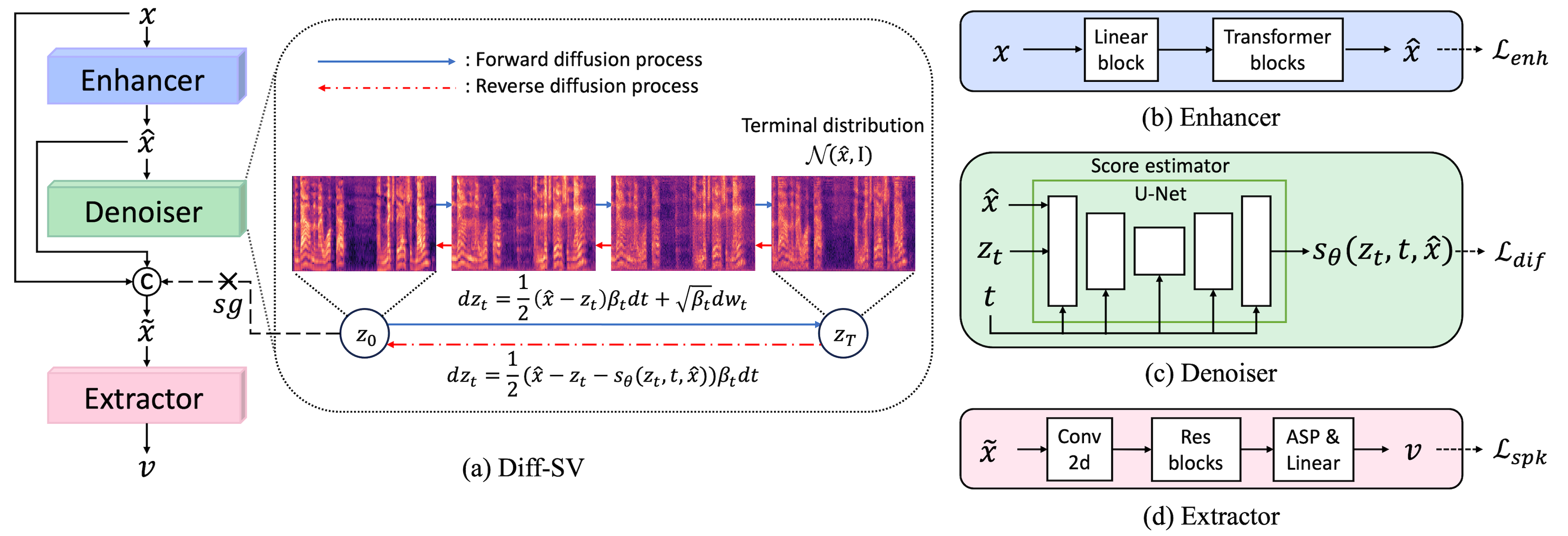}
    \vspace{-0.4cm}
    \caption{
        (a): Overview of Diff-SV. 
        Components of Diff-SV: an enhancer (b), a denoiser (c), and an extractor (d). 
        The extractor derives an embedding $v$ from $\tilde{x}$ concatenated with the original feature $x$, the enhanced feature $\hat{x}$, and the denoised feature $z_0$. 
        \textit{sg} indicates a stop gradient calculation operation. 
    }
    \label{fig:overview}
    \vspace{-0.2cm}
\end{figure*}

\section{Score-based Diffusion probabilistic models}
\label{sec:score}
DPMs \cite{sohl2015deep} have demonstrated promising results in the field of generative modeling by training the model to reverse data from noise. 
Recently, Song et al. \cite{song2020score} introduced a generalized framework for DPMs, employing SDEs and score, the gradient of the probability density function ($p$) for the data $z$. 
The score can be simply expressed as follows: 
\begin{equation}
\nabla_z \log p(z).
\end{equation}

Considering $z_0 \sim p_0$ and $z_T \sim p_T$ as the data and prior probability distributions, respectively, the forward diffusion process with a continuous time variable $t \in [0, T]$ is formulated as a solution to the following SDE: 
\begin{equation}
    \mathrm{d}z_t = f(z_t, t) \, \mathrm{d}t + g(t) \, \mathrm{d}w_t,
\end{equation}
where $w_t$ represents the standard Wiener process, also known as Brownian motion. 
The drift coefficient for $z_t$ and $t$ is denoted as $f(z_t, t)$, and $g(t)$ signifies the diffusion coefficient.
The infinitesimal timesteps close to $0$ are designated as $dt$. 
From $z_T$, the sample $z_0$ can be recovered via the reverse diffusion process (i.e., data sampling or data generation) as follows: 
\begin{equation}
\mathrm{d}z_t = \left[f(z_t, t)- g(t)^{2} \nabla_{z_t} \log p_t(z_t) \right]\mathrm{d}t + g(t) \, \mathrm{d}\bar{w},
\end{equation} 
In this equation, $\bar{w}$ indicates a standard Wiener process, with time changing from $T$ to $0$. 
As accurately computing $\nabla_{z_t} \log p_t(z_t)$ is difficult during the reverse process, score-based DPMs train a score estimator $s_\theta(z_t,t)$ to predict the score (i.e., score matching) for data sampling.

\begin{table*}[t!]
\caption{
    Experiment results (EER \%, $C_{det}^{min}$) obtained using the VoxCeleb1 test set and the noise scenarios synthesized using the MUSAN corpus under various SNRs ( $^{\dagger}$ : drawn from \cite{wu2021joint}). 
}
\centering
\label{table:in-in}
\resizebox{\textwidth}{!}{
\begin{tabular}{c|c|m{1.37cm}m{2.49cm}m{2.37cm}m{2.21cm}m{2.21cm}m{2.25cm}m{1.37cm}} \Xhline{2\arrayrulewidth}
Noise type & SNR  &\hfil Baseline    &\hfil VoiceID(2019)\cite{shon2019voiceid} $^{\dagger}$ &\hfil Wu \textit{et al.}(2021) \cite{wu2021joint} &\hfil NDML(2023) \cite{sun2023noise}  &\hfil Cai \textit{et al.}(2020) \cite{cai2020within} & \hfil ExU-Net(2022) \cite{DBLP:conf/interspeech/KimHSY22}& \hfil Diff-SV \\ \Xhline{2\arrayrulewidth}
\multicolumn{2}{c|}{\# Parameters}                    &\hfil 2.1M          &\hfil -         &\hfil -          &\hfil -        &\hfil -            &\hfil 4.81M        &\hfil 3.77M \\ \hline
\multicolumn{2}{c|}{Original test set}                &\hfil 2.37          &\hfil 6.79      &\hfil 7.6        &\hfil 2.90     &\hfil 3.12         &\hfil 2.76         &\hfil \textbf{2.35} \\\hline
\multirow{5}{*}{Babble} & 0                           &\hfil 11.56         &\hfil 37.96     &\hfil 20.11      &\hfil 10.96    &\hfil 11.78        &\hfil 9.57         &\hfil \textbf{8.74} \\  
                        & 5                           &\hfil 5.11          &\hfil 27.12     &\hfil 12.02      &\hfil 6.13     &\hfil 5.97         &\hfil 5.52         &\hfil \textbf{4.51} \\ 
                        & 10                          &\hfil 3.50          &\hfil 16.66     &\hfil 9.63       &\hfil 4.28     &\hfil 4.44         &\hfil 4.06         &\hfil \textbf{3.33} \\ 
                        & 15                          &\hfil 2.92          &\hfil 11.25     &\hfil 8.48       &\hfil 3.52     &\hfil 3.73         &\hfil 3.28         &\hfil \textbf{2.82} \\ 
                        & 20                          &\hfil 2.67          &\hfil 8.99      &\hfil 7.99       &\hfil 3.21     &\hfil 3.36         &\hfil 2.99         &\hfil \textbf{2.61} \\ \hline
\multirow{5}{*}{Music}  & 0                           &\hfil 7.21          &\hfil 16.24     &\hfil 12.92      &\hfil 10.84    &\hfil 7.79         &\hfil 7.35         &\hfil \textbf{6.04} \\ 
                        & 5                           &\hfil 4.47          &\hfil 11.44     &\hfil 10.1       &\hfil 6.52     &\hfil 5.23         &\hfil 4.9          &\hfil \textbf{3.96}  \\ 
                        & 10                          &\hfil 3.35          &\hfil 9.13      &\hfil 8.95       &\hfil 4.66     &\hfil 4.11         &\hfil 3.69         &\hfil \textbf{3.10} \\ 
                        & 15                          &\hfil 2.80          &\hfil 8.10      &\hfil 8.35       &\hfil 3.67     &\hfil 3.63         &\hfil 3.14         &\hfil \textbf{2.75} \\ 
                        & 20                          &\hfil 2.66          &\hfil 7.48      &\hfil 7.95       &\hfil 3.21     &\hfil 3.30         &\hfil 2.93         &\hfil \textbf{2.60} \\ \hline
\multirow{5}{*}{Noise}  & 0                           &\hfil 6.87          &\hfil 16.56     &\hfil 13.12      &\hfil 10.24    &\hfil 7.34         &\hfil 6.8          &\hfil \textbf{6.01}  \\ 
                        & 5                           &\hfil 4.82          &\hfil 12.26     &\hfil 10.57      &\hfil 6.96     &\hfil 5.65         &\hfil 5.23         &\hfil \textbf{4.52} \\ 
                        & 10                          &\hfil 3.81          &\hfil 9.86      &\hfil 9.28       &\hfil 5.02     &\hfil 4.35         &\hfil 4.07         &\hfil \textbf{3.49} \\ 
                        & 15                          &\hfil 3.02          &\hfil 8.69      &\hfil 8.59       &\hfil 3.91     &\hfil 3.85         &\hfil 3.39         &\hfil \textbf{2.93} \\ 
                        & 20                          &\hfil 2.71          &\hfil 7.83      &\hfil 8.1        &\hfil 3.40     &\hfil 3.44         &\hfil 3.1          &\hfil \textbf{2.64}  \\ \Xhline{2\arrayrulewidth}
\multicolumn{2}{c|}{Average (EER / $C_{det}^{min}$)}  &\hfil 4.37 / 0.237  &\hfil 13.52 / - &\hfil 10.24 / -  &\hfil 5.59 / - &\hfil 5.07 / 0.563 &\hfil 4.55 / 0.254 &\hfil \textbf{3.90} / \textbf{0.213}  \\ \Xhline{2\arrayrulewidth}
\end{tabular}
}
\vspace{-0.2cm}
\end{table*}

\section{Proposed Framework}
\label{sec:proposed}
In this section, we introduce Diff-SV, the proposed noise-robust SV framework that uses a score-based DPM, and the overview is illustrated in Fig. \ref{fig:overview} (a). 
The architecture of Diff-SV includes three primary components, namely, an enhancer, a denoiser, and an extractor. 
Initially, the enhancer pre-processes the original Mel-spectrogram by reducing noise. 
Subsequently, the DPM-based denoiser further refines the enhanced feature. 
Lastly, the extractor produces speaker embeddings by hierarchically processing features obtained from previous stages. 
All modules are trained in a unified approach.

\subsection{Enhancer}
Lu et al. \cite{lu2021study} employed a DPM-based speech enhancement system that uses the original noisy speech as the initial value in the reverse diffusion process to produce a denoised speech. 
However, we hypothesized that preliminary enhanced features can closely resemble to clean speech, and thus, allowing the DPM to sample denoised features more reliably compared to using the original noisy input. 
This assumption was similarly applied to the speech synthesis domain using DPM, which demonstrated impressive naturalness of the generated speech \cite{popov2021grad}. 
Therefore, Diff-SV utilizes an enhancer to obtain enhanced features for initializing the sampling process conducted in the denoiser, a DPM-based speech enhancement system. 

As depicted in Fig. \ref{fig:overview} (b), the enhancer ($f_{enh}$) comprises a linear block containing two fully connected layers, a Mish activation \cite{DBLP:conf/bmvc/Misra20}, a dropout layer, and transformer blocks \cite{vaswani2017attention}. 
The goal of the enhancer is to extract the enhanced features ($\hat{x}$) from the noisy Mel-spectrogram ($x$) as follows: 
\begin{equation}
    \hat{x} = f_{enh}(x),  \quad x,\ \hat{x} \in \mathbb{R}^{L \times F \times 1 \times B},
\end{equation}
where $L$, $F$, and $B$ represent the feature length, feature frequency, and batch size, respectively. 
The enhancer is optimized to minimize the L2 distance between the output and the clean Mel-spectrogram ($y$). %, similar to a denoising autoencoder. 
\begin{equation}
    \mathcal{L}_\text{enh} =  \frac{1}{B}\sum^B_{i=1} ||y_i - \hat{x}_i||^2_2.
\end{equation}

\subsection{Denoiser}
The denoiser, which comprises a score-based DPM, removes any residual noise from the enhanced features. 
We designed the denoiser by transforming all data distributions of infinite-time-horizon forward diffusion to $N(\hat{x}, I)$, rather than $N(0, I)$. 
This concept is inspired by the methods reported in \cite{popov2021grad}, which generalizes the data distribution of diffusion processes. 
Therefore, the terminal condition $z_T$ can be considered as data sampled from a normal distribution with the enhancer's output ($\hat{x}$) as the mean. 
We redefined the standard forward diffusion process (Equation (2)) as the following SDE: 
\begin{equation}
\mathrm{d}z_t = \frac{1}{2} (\hat{x} - z_t) \beta_t \mathrm{d}t + \sqrt{\beta_t} \mathrm{d}w_t,
\end{equation} 
where $\beta_t$ is the noise scheduling function \cite{song2020score}. 
Consequently, the denoiser can yield the denoised feature from the enhanced feature using the reverse diffusion process. 
To generate high-fidelity features for SV tasks, we employed ODEs with the random Wiener terms removed and formulated the sampling process as follows: 
\begin{equation} 
\begin{aligned}
&\mathrm{d}z_t = \frac{1}{2}(\hat{x} - z_t - s_{\theta}(z_t, \ t, \ \hat{x}))\beta_t \mathrm{d}t, &\\
\end{aligned}
\end{equation}
As $\nabla_{z_t} \log p_t(z_t)$ cannot be computed accurately in the reverse process, we estimated the score using $s_\theta$. 
As shown in Fig. \ref{fig:overview} (c), the structure of the score estimator is based on U-Net as reported in \cite{popov2021grad} and takes $z_t$, $t$, and $\hat{x}$ as input to predict the score corresponding to each time point $t$. 
Thus, the feature $z_0$, which is deterministically derived from $z_T$, can be used in the SV framework owing to less perturbation of the speaker distribution and fast sampling. 

To train $s_\theta$, we calculated the expectation by marginalizing over a tractable transition kernel. 
Given that $p(z_t | z_0)$ follows a Gaussian distribution, the loss function for score estimation is as follows:
\begin{equation}
\mathcal{L}_\text{dif} = \mathbb{E}_{t \sim \mathcal{U}(0, T)} \mathbb{E}_{z_0 \sim p_0} \mathbb{E}_{\epsilon_t \sim \mathcal{N}(0, I)} \lVert s_{\theta}(z_t, \ t, \ \hat{x}) - \sigma_t^{-1} \epsilon_t \rVert_2^2,
\end{equation}
where $\sigma_t = \sqrt{1 - e^{-\int_0^t \beta(s) ds}}$.

\subsection{Extractor}
In the proposed framework, the ResNet-based extractor ($f_{ext}$) derives embedding ($v$) (Fig. \ref{fig:overview} (d)). 
To configure a unified noise-robust SV framework, the original, enhanced, and denoised features are jointly fed to the extractor. 
However, we empirically observed that the output of Diff-SV failed to converge, possibly due to gradient exploding or vanishing problems caused by repeated Gaussian denoising operations during the denoiser's reverse process. 
Therefore, we delivered $z_0$ after the stop gradient calculation operation (\textit{sg}) to prevent the exploding or vanishing gradient of the denoiser from occurring by backpropagation through the extractor's objective function as follows: 
\begin{equation}
\begin{aligned}
    &\qquad \qquad v = f_{ext}(\tilde{x}),& \\
    &\tilde{x} = [x, \ \hat{x}, \ \textit{sg}(z_0)],\ \tilde{x} \in \mathbb{R}^{L \times F \times 3 \times B}, &
\end{aligned}
\end{equation}
where $[\cdot]$ denotes the concatenation of each element on the channel axis. 
Thus, using a combination of (\textbf{\lowercase\expandafter{\romannumeral1}}) noisy but non-destructive original features $x$, (\textbf{\lowercase\expandafter{\romannumeral2}}) enhanced features $\hat{x}$ considering the target task, and (\textbf{\lowercase\expandafter{\romannumeral3}}) independently denoised features $z_0$, which are almost similar to clean features, the extractor can enrich the information from different perspectives while mitigating the distortion of speaker information via speech enhancement. 
The embeddings were optimized to classify the speakers included in the training data using an additional linear layer ($W$), and we employed the additive angular margin (AAM)-Softmax function \cite{deng2019arcface}. 
\begin{equation}
\mathcal{L}_{spk} = -\frac{1}{B}\sum^{B}_{i=1}\log\frac{e^{s \cdot (cos(\theta_{c_{i}, i} + m))}}{e^{s \cdot (cos(\theta_{c_{i}, i} + m))} + \sum_{j \not= y_i}e^{s \cdot (cos(\theta_{j, i}))}} , 
\end{equation}
where $\theta_{j, i}$ is the angle between the $j$-th weight vector $W_j$ and $i$-th embedding vector $v_i$, and $c_i$ denotes the speaker label of $v_i$. 
Additionally, the scaling factor ($s$) and margin ($m$) are set to 0.3 and 30, respectively. 

Finally, Diff-SV was trained to optimize the following three losses: 
\begin{equation}
    \mathcal{L} = \mathcal{L}_\text{enh} + \mathcal{L}_\text{dif} + \mathcal{L}_\text{spk}.
\end{equation}

\section{Experiments}
\label{sec:exp}
\subsection{Datasets}
\label{sec:datasets}
The models were trained using VoxCeleb1 \cite{DBLP:conf/interspeech/NagraniCZ17} training data, consisting of 1,211 speakers. 
We employed MUSAN corpus \cite{snyder2015musan} to generate noise data, which we divided into non-overlapping training and test subsets. 
Noisy data for training was constructed using the MUSAN training subset with a randomly selected signal-to-noise ratio (SNR) between 0--20, further augmented with room impulse response reverberation, pitch shift, and gain variations. 
We constructed three test conditions to evaluate the noise robustness of our proposed model from different perspectives. 
First, \textit{in-domain speech evaluation with in-domain noisy data} was conducted by augmenting the VoxCeleb1 test set with the SNR values \ \{0, 5, 10, 15, and 20\} for each noise type in the MUSAN test subset. 
Although the noise sources are separated, the data distributions within the same corpus could be similar. 
Therefore, we organized an \textit{in-domain speech evaluation with out-of-domain noisy data} using a separate noise source.   
We used the Nonspeech100 dataset \cite{hu2010tandem} as the out-of-domain noise source and evaluated the system with the same configurations as those corresponding to the in-domain noise evaluation conditions. 
Finally, to verify the generalizability of the system, we conducted \textit{out-of-domain speech evaluation with out-of-domain noisy data}. 
To achieve this, we used the VOiCES development and evaluation dataset. 
The VOiCES dataset \cite{richey18_interspeech} was collected at different distances and under various acoustic conditions using array microphones in rooms of different sizes. 
We evaluated the models based on the parameter that achieved the best performance in the clean VoxCeleb1 test scenario. 

\subsection{Implementation details} 
We inputted an 80-dimensional log Mel-spectrogram extracted by conducting a 512-point fast Fourier transform with a Hamming-window width of 25 ms and 10-ms hopping. 
Diff-SV was trained using the AMSGrad optimizer \cite{reddi2018convergence} with a mini-batch of 160. 
The initial learning rate (LR) was 1e-3, which was decreased to 1e-7 over four cycles for 320 epochs using a cosine LR scheduler. 
The models were compared based on their equal error rates (EER) obtained using the cosine similarity score and the minimum detection cost function ($C_{det}^{min}$). 
The enhancer in the proposed framework consists of four transformer blocks with a hidden size of 80 dimensions, and the architecture of our extractor is identical to ResNet structures reported in \cite{DBLP:conf/interspeech/KimHSY22}, respectively. 
The baseline had the same structure as that of the extractor, except that the channel size of the first convolution layer was changed from 3 to 1 using only the original features as the input. 
Additional details are present under the experimental code at \url{https://github.com/wngh1187/Diff-SV}. 

\section{Results and discussion}
\label{sec:results}
Table \ref{table:in-in} lists the in-domain evaluation results using the VoxCeleb1 test set and the MUSAN evaluation partition. 
The baseline surpasses the average results obtained from noise-robust SV systems developed in recent years
This improvement can be attributed to the optimization of the advanced objective function (AAM-Softmax) and the application of various data augmentation techniques during training. 
Nevertheless, the baseline performance deteriorates under considerably noisy conditions (e.g., under babble and noise with an SNR of 0 dB), indicating the need for specialized noise-reduction approaches. 
Our proposed method, Diff-SV, demonstrates enhanced noise robustness with respect to that of the baseline across all evaluation scenarios. 
Diff-SV achieves a relative error reduction (RER) that is 14.29\% higher while using fewer parameters than ExU-Net, the top-performing noise-robust system (4.55\% vs. 3.9\%). 

Table \ref{table:in-out} presents the results of each model under out-of-domain noise evaluation with in-domain speech conditions. 
In contrast to in-domain noise evaluation results, ExU-Net showed superior performance with respect to the baseline, indicating the effectiveness of the noise compensation. 
Moreover, Diff-SV outperforms all models, achieving an average EER and $C_{det}^{min}$ of 4.65\% and 0.247, respectively. 
These results reveal the noise robustness of the proposed model under in- and out-of-domain noise scenarios.

\begin{table}[]
%\vspace{0.1cm}
\caption{
    Experimental results (EER \% and $C_{det}^{min}$) obtained on the VoxCeleb1 test set synthesized with an out-of-domain noise source (Nonspeech100). 
}
\centering
\label{table:in-out}
\resizebox{\linewidth}{!}{
\begin{tabular}{c|c|c c c c} \Xhline{2\arrayrulewidth}
Noise type & SNR                                      & Baseline     & NDML      & ExU-Net       & Diff-SV \\ \Xhline{2\arrayrulewidth}
\multirow{5}{*}{Nonspeech} & 0                        & 10.25        & 20.49     & 8.39          & \textbf{8.23} \\
                           & 5                        & 6.27         & 15.09     & 5.59          & \textbf{5.06} \\
                           & 10                       & 4.28         & 11.96     & 4.36          & \textbf{3.85} \\
                           & 15                       & 3.32         & 9.96      & 3.74          & \textbf{3.19} \\
                           & 20                       & 2.99         & 8.64      & 3.29          & \textbf{2.89} \\ \Xhline{2\arrayrulewidth}
\multicolumn{2}{c|}{Average (EER / $C_{det}^{min}$)}  & 5.42 / 0.286 & 13.23 / - &  5.08 / 0.286 & \textbf{4.65} / \textbf{0.247} \\ \Xhline{2\arrayrulewidth}
\end{tabular}
 }
\vspace{-0.2cm}
\end{table}

Table \ref{table:out-out} displays the results obtained from each model using the VOiCES development and evaluation datasets. 
Diff-SV achieves RERs of 15.38\% and 24.98\% for each dataset with respect to the baseline due to the proposed noise-compensation modules. 
Based on the evaluation results obtained from the out-of-domain speech and noise datasets, we confirmed the outstanding generalizability of our proposed framework. 

\begin{table}[]
\vspace{0.1cm}
\caption{Results of EER and $C_{det}^{min}$ on the VOiCES development and evaluation datasets.}
\centering
\label{table:out-out}
\resizebox{\linewidth}{!}{
\begin{tabular}{c | c c c c c}
\Xhline{2\arrayrulewidth}
& Baseline & VoiceID & Wu \textit{et al.} & ExU-Net & Diff-SV \\ \hline
Dev & \hfil 2.47 / 0.145  & \hfil 13.57 / -  & \hfil 12.51 / - & \hfil 6.53 / 0.381 & \hfil \textbf{2.09} / \textbf{0.126}\\ 
Eval & \hfil 11.45 / 0.438 & \hfil - / - & - / - & 23.32 / 0.851 & \hfil \textbf{8.59} / \textbf{0.358} \\ 
\Xhline{2\arrayrulewidth}
\end{tabular}
}
\vspace{-0.1cm}
\end{table}

An ablation study was performed to evaluate the efficacy of each component of the proposed framework. 
The models were evaluated based on their corresponding EER and $C_{det}^{min}$ under the same evaluation scheme of Table \ref{table:in-in}. 
The results are displayed in Table \ref{table:ablation}, in which system \#1 corresponds to Diff-SV. 
System \#2 signifies an architecture that employed a pre-trained enhancer and denoiser to train an extractor, rather than applying unified training. 
This approach is sub-optimal because the enhancer and denoiser are not specifically designed for SV tasks, making early stopping challenging. 
In addition, feeding only the denoised features into the extractor without applying the \textit{sg} operation (system \#3), instated of following the hierarchical input structure, leads to non-convergence. 
We suspect that this stems from the gradient vanishing or exploding in the repeated score matching operation during the reverse process of the denoiser. 
Furthermore, the results of systems \#4 and \#5 reveal that excluding the denoiser and enhancement loss significantly degrades the generalization performance. 
Therefore, these results emphasize the importance of integrating the components of Diff-SV with a hierarchical input structure.

\begin{table}[]
\caption{
    Ablation experiments conducted on Diff-SV. 
    N/C stands for non-convergence. 
}
\centering
\label{table:ablation}
\resizebox{\linewidth}{!}{
\begin{tabular}{l|c|c c c c}
\Xhline{2\arrayrulewidth}
System configuration & \# Params  & \multicolumn{2}{c}{Original} & \multicolumn{2}{c}{Noises Average} \\ \hline
\#1 Diff-SV                                  & 3.77M & \multicolumn{2}{c}{ \textbf{2.35} / \textbf{0.141}} & \multicolumn{2}{c}{\textbf{4.00} / \textbf{0.213}} \\
\#2 \ \; \textit{w/o} unified training         & 3.77M & \multicolumn{2}{c}{ 2.46 / 0.151 } & \multicolumn{2}{c}{ 4.22 / 0.220 } \\ 
\#3 \ \; \textit{w/o} hierarchical structure & 3.76M &\multicolumn{2}{c}{ N/C } & \multicolumn{2}{c}{ N/C } \\
\#4 \ \; \textit{w/o} denoiser               & 3.20M  & \multicolumn{2}{c}{ 2.52 / 0.152 } & \multicolumn{2}{c}{ 4.53 / 0.234} \\
\#5 \quad \; \textit{w/o} $\mathcal{L}_\text{enh}$           & 3.20M & \multicolumn{2}{c}{ 3.35 / 0.178 }  & \multicolumn{2}{c}{ 5.99 / 0.300 } \\
\Xhline{2\arrayrulewidth}
\end{tabular}
}
\vspace{-0.1cm}
\end{table}

\begin{figure}[]
  \begin{center}
    \centering
    \includegraphics[width=\linewidth]{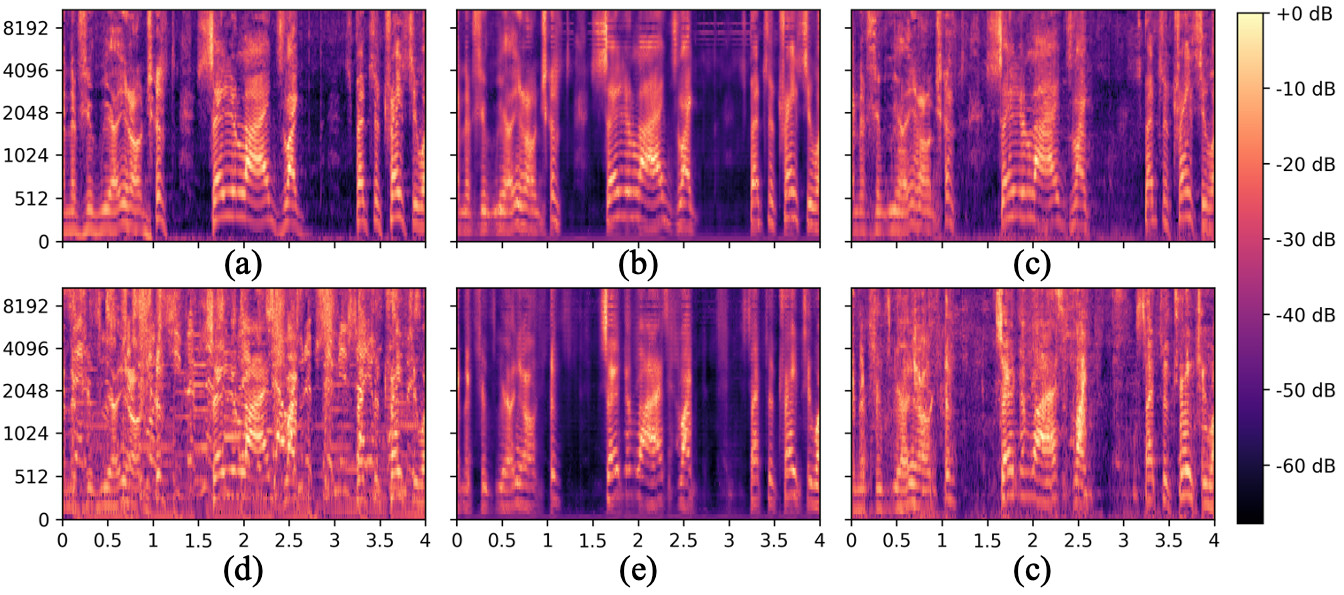} 
    \vspace{-0.4cm}
    \caption{Visualization of features inside the Diff-SV framework. 
    (a) and (d): Spectrograms of the original test utterance and the augmented test utterance with music (SNR 0). 
    (b) and (e): Enhancer outputs for inputs (a) and (d). 
    (c) and (f): Denoiser outputs for inputs (b) and (e).
    The x- and y-axes of each spectrogram represent time and frequency, respectively. 
    }
    \label{fig:spec}
  \end{center}
  \vspace{-0.1cm}
\end{figure}

We visualized the internal spectrograms of the Diff-SV framework to verify the denoising effectiveness. 
Fig. \ref{fig:spec} presents the original spectrogram (a) of a randomly chosen speech from the VoxCeleb1 evaluation set, a noisy spectrogram (d) synthesized by adding music at an SNR of 0 dB, and the output features of the enhancer (b), (e) and denoiser (c), (f). 
As noise is introduced, the original utterance becomes contaminated with signals of varying shapes and frequency bands ((a) vs. (d)). 
Although an enhancer trained to map input to clean speech can effectively eliminate noise, simultaneous optimization for the target task results in over-smoothing (blurring) or collapse (horizontal lines) of low- and high-frequency information ((a) and (d) vs. (b) and (e)). 
Due to the difficulty of accurately removing Gaussian noise during the inversion process, the output of the denoiser retains more noise but exhibits better pitch information owing to the superior generalizability of DPM ((b) and (e) vs. (c) and (f)). 
Therefore, by simultaneously supplying features with complementary properties, Diff-SV is capable of producing speaker embeddings that are highly discriminative and noise-tolerant. 

\section{Conclusion}
Based on the impressive generative capabilities of DPM, we propose a unified noise-robust SV framework that uses a DPM-based speech enhancement system. 
Using the ODE solver of the score-based DPM, Diff-SV generates denoised features from the enhancer's output in an efficient and deterministic manner. 
Furthermore, the complementary hierarchical inputs improve the extractor’s ability to derive discriminative and noise-robust speaker embeddings. 
Our proposed model outperforms recently reported models, including baselines, under both in- and out-of-domain scenarios. 
Additionally, we demonstrate the effectiveness of Diff-SV and the importance of each component through visualization and ablation analyses. 
We intend to explore the incorporation of other generative approaches to improve noise compensation capabilities in future studies. 

\vfill\pagebreak

\bibliographystyle{IEEEbib}
\bibliography{strings,refs}

\end{document}